# Secondary Concentration Plateau and Formation of Flow Stagnation by Ion Concentration Polarization in Microchannels

Junsuk Kim[1], Keon Huh[1], Ali Mani[2*] and Sung Jae Kim[1,3,4*]

[1]*Department of Electrical and Computer Engineering,*
*Seoul National University, Seoul, 08826, Republic of Korea*
[2]*Department of Mechanical Engineering,*
*Stanford University, Stanford, 94305, USA*
[3]*Nano Systems Institute, Seoul National University, Seoul, 08826, South Korea*
[4]*Inter-university Semiconductor Research Center,*
*Seoul National University, Seoul, 08826, South Korea*

[*]Correspondence should be addressed to Prof. Ali Mani and Prof. Sung Jae Kim
E-mail: (Ali Mani) alimani@stanford.edu, (SJKim) gates@snu.ac.kr

# Abstract

Ion transport in perm-selective media has been extensively investigated in recent years owing to its applications in advancements of fundamental understanding of nanoscale electrokinetics as well as innovative engineering applications. A key phenomenon in this context is ion concentration polarization (ICP) that occurs near perm-selective nanoporous membranes or nanochannels under dc bias. In classical settings, involving voltage-driven transport of ions through perm-selective membranes, concentration polarization is well understood as formation of steep concentration drop to reach near zero concentration (depletion layer) at the anodic side of cation-selective membranes. In contrast to this classical description, we demonstrate here that when ICP is driven in a microchannel and coupled with electroosmotic flows a long-tailed fluorescent layer in front of diffusion layer is formed, which can be characterized as an additional plateau in the concentration profile. Using a micro/nanofluidic platform, *in situ* visualizations, and local concentration measurements by microelectrode arrays we quantify that this plateau layer has a concentration of 60 % of bulk electrolyte concentration, regardless of the bulk concentration itself. Using a mathematical analysis of conservation laws for flow, charge, and salt, we demonstrate that this plateau region is related to the formation of a flow stagnation point due to the competition between electro-osmotic and induced pressure driven flows. Consistent with the experiments, this analysis predicts the plateau concentration to be 60% of the bulk electrolyte. Thus, this plateau concentration region completes the picture of ICP in microchannels which has drawn significant attention from the fields of electrokinetics and micro/nanofluidics.

Ion concentration polarization (ICP) is a fundamental electrokinetic phenomenon that occurs near perm-selective nanoporous membranes or nanochannels[1-3]. ICP involves significant concentration gradient that is induced by an applied dc bias across perm-selective membranes. The concentration gradient regions are called ion depletion zone (low concentration) and ion enrichment zone (high concentration) which are respectively formed at the anodic and cathodic sides in the case of cation-selective membranes (*e.g.* Nafion) and *vice versa*[4]. However, the complete scenario of ICP phenomenon is still lacking when this phenomenon occurs in a microchannel, because of additional complexities including electroconvective effects[3,5-11], geometrical confinement[12-14], surface conduction[15-18], electro-osmotic flow[7,16,19], and coupling with other physical effects associated with wide range of physical scale ranging from nanometers to centimeters[20-22]. Because of this complexity, the experimental and theoretical investigations have often been focused on the componential problems of ICP hindering an in-depth understanding of the fully coupled problem.

One of the key characteristics of ICP is that charged particles are being repelled from the ion depletion zone which acted as a virtual barrier due to the balance of hydrodynamic drag force and electrophoretic force on the charged molecules[20,23-28] mainly depending on the concentration profile of the ICP layer. Thus, understanding of concentration profiles[29,30] not only advances fundamental science in the field of electrokinetics but it also guides development of novel future engineering applications. Classically, the concentration profile of diffusion layer was predicted to be linear in the early ICP model as shown in Figure 1(a). The model only considered diffusion and electro-migration as the mechanism of ion transport based on the Nernst-Planck equation[1]. Only Ohmic and limiting ionic current could be explained with this early model. With more theoretical and experimental studies, convection effects were found to play an important role in physical settings when currents beyond the limiting values were measured. Strong vortices generated by nonlinear electrokinetic flows efficiently mix the electrolyte in the depletion zone causing the concentration of depletion zone to be nearly flat and below 10 % of bulk concentration as shown in Figure 1(b)[3,5,31,32]. These effects have led to identification of various mechanisms of overlimiting current, which is followed by Ohmic and limiting current regimes. These models were further improved by considering generation of secondary vortices (Figure 1(c)). Step-wise concentration profile in the ICP layer was verified with an in-depth experimental investigation of multiple vortical instabilities inside the ICP layer[29,33]. These models have been employed to describe the core mechanism of biomolecular preconcentrators[20,23-28] (due to sharp concentration gradient) and desalination / purification devices[34-36] (due to low concentration compare to the bulk concentration ($< 0.1$ $c_0$)). However, repeated experiments have shown anomalous phenomena of unexpected long-tailed fluorescent signal followed by sharp preconcentrated plug, which has remained unexplained.

This work presents an in-depth investigation of the spatiotemporal concentration profiles outside of depletion zones in ICP, which is intermediate region between bulk and ion depletion layer. Unlike the diffusion-convection layer which has sudden drop in concentration followed by near zero concentration (Figure 1(c)), we present measurement data demonstrating an additional concentration plateau when ICP occurs in a microchannel (Figure 1(d)). In the geometrical setting considered here, this propagating plateau reaches a concentration of nearly 60 % of bulk concentration as confirmed by an *in-situ* visualization with/without fluorescent dye and an *in-situ* conductivity measurement using microelectrode array. Furthermore, we present a simple mathematical analysis revealing that the secondary concentration plateau is formed due to formation of a stagnation flow as result of competition between electro-osmotic and induced pressure driven flow in the microchannel, which is also found to be independent of bulk concentration.

Figure 2(a) and 2(b) show the snapshot and schematic of fabricated device, which had patterned Au microelectrode array on the glass substrate to measure the local electric potentials along the microchannel (width 200 μm, depth 15 μm and length 1 cm). Note that the nanojunction was patterned at the center of the microchannel. See the supporting note 1 for detailed fabrication procedures of the device. Figure 2(c) shows the schematics of the operating platform of the *in-situ* visualization and concentration profiling experiments. See supporting note 2 for full details of experimental setup. Electrolyte solution of KCl (0.1 mM ~ 100 mM, Sigma-Aldrich, USA) mixed with 20 μM fluorescent dye (Alexa 488, Invitrogen, USA) and carboxylate-modified FluoSpheres ($d$ = 1 μm, yellow-green fluorescent, Invitrogen, USA) was injected into the main microchannel. Also, KCl electrolyte (10 mM ~ 1 M) at least 10 times higher than main microchannel was injected to the buffer microchannel which was mixed with 20 μM Alexa 488. The reasons for using high KCl concentration at the buffer microchannel were to minimize the effects of concentration change of ion enrichment zone and to minimize voltage drop in the buffer solution[24,37]. Electrical voltage was applied with source measure unit (Keithley 236, USA) while current-time responses were obtained by Ag/AgCl electrodes simultaneously. The applied voltage was chosen in the regime of overlimiting regime[13] causing ICP phenomenon near the nanoporous membrane. Also, other four source measure units (Keithley 236, 237, 238, USA) measured the local potentials of the microchannel simultaneously with patterned Au microelectrode array. Since there were 7 patterned Au microelectrodes and 4 measuring units, only 4 local potentials were measured at each experiment. Then, the average conductivity of the electrolyte between each microelectrode pairs, σ, was calculated by the Ohm's law $\sigma = i/((|\Delta V|A)/\Delta L)$, where $i$ is the current passing through the microelectrode, $\Delta V$ is the voltage difference between microelectrodes, $A$ (14.5 μm × 200 μm) was the cross-sectional area of the microchannel and $\Delta L$ (250 μm or 500 μm) was the gap between the microelectrodes. The local concentration was estimated by reference conductivity of the KCl solutions. The reference conductivity in 0.01, 0.1, 1, 10, 100 mM KCl electrolyte were 1.61 μS cm$^{-1}$, 14.5 μS cm$^{-1}$, 150 μS cm$^{-1}$, 1.43 mS cm$^{-1}$, 13.0 mS cm$^{-1}$ respectively[29]. Based on these sets of conductivity values, the measured local average conductivities were converted into local average concentration values. During the electrode measurement, there are several side effects which are carefully remedied such as faradaic reaction[38,39], induced charge electrokinetic effects[40,41], Joule heating[42] as we shall explain in supporting note 3.

Figure 3(a) shows the time evolving snapshots at the main microchannel with the applied voltage of $+V_{app}$ = 70 V. The main microchannel was filled with 10 mM KCl ($c_0$) as a background electrolyte, while buffer microchannel was filled with 1 M KCl in this experiment. The ion depletion zone was formed next to the nanoporous membrane and fluorescent intensity increased due to the preconcentrated fluorescent dyes with the applied voltage. Four microelectrodes with odd-numbered position which had 500 μm gap between electrodes were used to measure the local potentials of the solution.

The intervals of interest are indicated by the dotted arrows in Figure 3(a), which indicates an unexpected long-tailed fluorescent signal propagating to the left of the maximum intensity of sharp plug. $c_{jk}$ indicated the average concentration between $j$th and $k$th microelectrode. The average spatiotemporal concentrations between the electrodes were shown in Figure 3(b) with red ($c_{13}$), yellow ($c_{35}$) and green ($c_{57}$) circles. The measurement zones are shown by rectangles with the same color in Figure 3(a). We next describe the transitions observed in the red rectangular ($c_{13}$) region. Between 0 ~ 1 minute, fluorescent intensity remained the same as the initial condition of mixed solution at 10 mM (see red circles of 0 ~ 1 min in Figure 3(b)). Between 1 ~ 2 minutes, tailed fluorescent intensity region propagated through this zone and, correlated with this observation, the local average concentration decreases to ~ 6 mM (see red circles of 1 ~ 2 min in Figure 3(b)). After 2 minutes, the fluorescent intensity of the tail remained fixed inside the zone indicated by the red rectangle and the local average concentration stayed fixed around 6 mM (see red circles of 2 ~ 5 min in Figure 3(b)). This description can also be applied to other regions; yellow ($c_{35}$) and green ($c_{57}$) rectangular regions. As a result, the concentration of tailed fluorescent dye regions was verified to be ~ 0.6 $c_0$ based on the visualization and measured conductivity data. Both fluorescence signal and conductivity measurements indicate an unexpected long concentration plateau, which propagates toward the anode (increasing dotted arrow region in Figure 3(b)). To investigate the time-dependency, long-term measurement was also conducted as shown in the inset of Figure 3(b). Here, the nanoporous membrane was apart from the first microelectrode by a distance of 1500 μm. The concentration of tailed fluorescent region maintained around 0.6 $c_0$ up to 1 hour confirming that this secondary plateau concentration zone was not a temporal phenomenon. See the supporting video 1 which showed *in-situ* measured local concentrations as well as $i$ and $V_j$ ($j$ is the electrode number). Note that 1 μm FluoSpheres were being repelled and accumulated at the interface between bulk and the secondary plateau layer.

In addition, the time when $c_{13}$ (red circle) reached plateau concentration (~ 6 mM) was close to the time when $c_{35}$ (yellow circle) started to decrease ($t$ = ~ 2 min). Likewise, the time when $c_{35}$ (yellow circle) reached plateau concentration (~ 6 mM) was close to the time when $c_{57}$ (green circle) started to decrease ($t$ = ~ 3 min). From these observations, one can infer that the boundary between zones at bulk concentration ($c_0$) and plateau concentration (~ 0.6 $c_0$) was passing through the specific electrodes at those times.

Conventional fluorescence dye such as Rhodamine 6G and Alexa 488 had been widely used to visualize ICP near the permselective nanochannel or nanoporous membrane[24,43-45]. As mentioned earlier, the tailed fluorescent dye also indicated the plateau concentration region. To exclude the possibility of any influence of the fluorescent dye in formation of this region, the identical experiment without fluorescent dye was conducted. These experiments consistently showed that the plateau region appeared regardless of the dye. See supporting video 2 and note 4 for the experiment without fluorescent dye. Negligible role of fluorescent dye was attributed to the extremely low concentration compared to the background electrolyte concentration. In Figure 3, fluorescent dye concentration in plateau region was estimated as 90 μM (initial 20 μM amplified by 450 %) by image analysis, while the background ions concentration was measured to be 6 mM (0.6 × 10 mM). Based on the measured data, one can calculate the contribution of each ionic species to the overall current. Their contribution is proportional to the electrical mobility and concentrations so that the contribution of dye was calculated to be only 0.68% ($\mu_{Alexa488}c_{Alexa488}/\mu_{Cl}c_{Cl}$) compared to the anion ions. Here, the electrical mobility of Alexa488 and Cl$^-$ were $\mu_{Alexa488}$=36.0×10$^{-9}$ and $\mu_{Cl}$=79.1×10$^{-9}$ m$^2$/Vs, respectively[46]. Therefore, the effect of fluorescent dye was turned out to be negligible based on these observations mentioned above.

To explain the formation of this concentration plateau, we next present a simple mathematical analysis. The domain was decomposed by zone (1): bulk region, zone (2): the secondary plateau and zone (3): the ion depletion zone as shown in Figure 4(a). Average concentration normalized by $c_0$, $\bar{c}_{(i)}$, of each zone $i$ was $\bar{c}_{(1)} = 1$, $\bar{c}_{(3)} \ll 1$ [34] and $\bar{c}_{(2)} = \alpha$ which is measured to be 0.6, but in this analysis is yet unknown. We start by analyzing the velocity profiles. In this setup, the long microchannel has two sections of equal size, to the left and right of the nanofluidic element. The electric field is active only in the left half of the channel, but this field is responsible for driving the flow through the entire channel. Given that most of the left half is occupied by zone (1) which also contributes to half of hydraulic resistance of the system, we conclude that the net averaged velocity in the entire channel is $\bar{u} = U/2$ where $U = -\varepsilon\zeta|\mathbf{E}|/\mu$ is the electroosmotic velocity induced in zone (1) where $\varepsilon$ and $\mu$ are the permittivity and viscosity of water, respectively, $\zeta$ is the zeta potential of the channel wall, and $\mathbf{E}$ is the electric field in zone (1). In this notation over-line indicates the average of the velocity profile. The reason for factor (1/2) is that half of the electroosmotic flow is cancelled by an induced internal pressure to maintain the same velocity in both left- and right-sides of the microchannel. In zone (2), the net conductivity is smaller by a factor of $\alpha$, and thus by conservation of current, the electric field must be larger by a factor of $1/\alpha$. This will lead to a higher induced electroosmotic flow in zone (2), but also proportionally higher and opposite induced pressure-driven flow in order to maintain continuity of a constant $\bar{u}$ equal to that in zone (1). Considering $y$ as the normalized coordinate in the wall-normal direction, the velocity profile in zone (2) should be

$$u_{(2)}(y) = U\left[\frac{1}{\alpha} + 6\left(\frac{1}{2} - \frac{1}{\alpha}\right)y(1-y)\right]. \tag{1}$$

This flow field at the centerline, $y = 0.5$, results in a velocity $u_c$ (3/4) – 1/(2$\alpha$). In zone (1) the velocity profile is positively valued ($\alpha = 1$), but in zone (2) for $\alpha < (2/3)$, we have a backward flow at the centerline. This implies that there is a stagnation point at the centerline between zones (1) and (2). This stagnation point was also confirmed by an accumulation of micro-beads in supporting video 1 and video 2. The streamline that passes through the stagnation point (see the thick line in Figure 4(a), separates the fluid into two regions. The outer fluid moves from left to right and brings slat-rich electrolyte from bulk to zone (2). The inner region recirculates the depleted salt from zone (3). We assume that convection is sufficiently faster than diffusion in zone (2)[3,5,47], and thereby in zone (2) salt concentration in the region outside of the separating streamline ($c \approx \bar{c}_{(1)}$) is isolated from the salt inside of the separator streamline ($c \approx 0$). Given that $\overline{c}_{(2)} = \alpha \overline{c}_{(1)}$, the separator streamline must be at $y = \alpha/2$ in zone (2), in order to maintain the average conductance consistently. Applying continuity of the fluid flow between zone (1) and zone (2) for the region outside of the separator streamline, implies $U/2 = 2\int_{y=0}^{y=\alpha/2} u_{(2)}(y)dy$. Substitution for $u_{(2)}(y)$ from Equation (1) and solving the resulting equation leads to $\alpha = 2 - \sqrt{2} \approx 0.59$, which is very close to the 60 % value measured experimentally as is in fact within the experimental accuracy. As the analytical investigation indicated, the formation of the secondary plateau of 0.6 $c_0$ was free from initial bulk concentration ($c_0$). See supporting note 5 for detailed analytical derivation.

In the experiments we investigated bulk concentrations ranging from 0.1 mM to 100 mM. As shown in Figure 4(b), the secondary plateau layer formed in all cases resulted in plateau concentration nearly 60 % of the bulk concentration. However, the ratio slightly increased as the bulk concentration became higher. For example, the ratio in the case of 0.18 mM and 1.03 mM were 56.5% and 59%, respectively. See supporting note 6.

ICP process has been extensively studied in recent years since it plays a core role in innovative engineering applications such as biomolecular preconcentrator and small-scale electro desalination device. Consequently, understanding of fundamental characteristics of ICP has evolved as well, especially when it comes to quantitative description of the concentration profiles. In this work, the spatiotemporal concentration profile of diffusion-convection layer of ICP had been rigorously investigated using patterned microelectrode array in micro/nanofluidic platform. This was the first experimental verification that secondary concentration plateau region of 60 % of $c_0$ is formed in front of diffusion layer in common setups utilizing ICP, unlike most-recent diffusion-convection layers which shows sudden drop and maintaining near zero concentration. The propagating plateau concentration profile with nearly 60 % of $c_0$ was confirmed with *in situ* fluorescent dye visualization and *in situ* microelectrode conductivity measurements. To explain this anomalous concentration profile in general, a simple analytical investigation was carried out to confirm that the formation of this secondary plateau and its independence from the bulk concentration in the sample electrolyte. Our analysis revealed the competition between pressure-driven flow and electroosmotic flow as the key mechanism for the formation of the plateau region and showed that this region is separated from the standard diffusion layer by a zone involving flow stagnation. This secondary plateau concentration layer together with a stagnation point should be considered for the complete picture of ICP in microchannels and would play a key role not only for advancing novel micro- and nano-scale-electrokinetic functionalities but also for proper quantitative analysis of existing applications.

# Acknowledgements

This work is supported by Basic Research Laboratory Project (NRF-2018R1A4A1022513) and the Center for Integrated Smart Sensor funded as Global Frontier Project (CISS- 2011-0031870) by the Ministry of Science and ICT and Korean Health Technology RND project from the Ministry of Health and Welfare Republic of Korea (HI13C1468, HI14C0559). Also authors acknowledged the financial supports from BK21 Plus program of the Creative Research Engineer Development IT, Seoul National University. S. J. Kim acknowledged the financial support from LG Yonam Foundation, Korea.

# Additional Information

## Competing interests

The authors declare no competing interests (both financial and non-financial).

# Author Contributions

J. Kim conceived the observation, designed the micro/nanofluidic device and mainly conducted experiments, K. Huh conducted the experiment for bulk concentration effect. A. Mani provided analytical investigation. S. J. Kim supervised the project. All authors wrote the manuscript.

**FIGURE CAPTIONS**

**Figure 1.** Schematic diagrams of the concentration profile of ICP layer. (a) Early model of ICP layer only with diffusion and drift. (b) Nonlinear electrokinetic slip induced by strong vortex was added to (a). (c) Multiple vortex caused by primary vortex was added to (b). (d) Proposed model including plateau concentration next to outside of the ion depletion zone.

**Figure 2.** (a) Snapshot of the fabricated device with the patterned Au microelectrode array. The device was fabricated with conventional soft lithography method and lift-off process. (b) Schematic of the device with the applied voltages at each reservoir and patterned Au microelectrode array. (c) Diagram of operating platform with microscopic view of the fabricated device.

**Figure 3.** (a) Microscopic view of time-evolving snapshots at the main microchannel with 10 mM KCl ($c_0$) as a background electrolyte. Tailed fluorescent dye was being propagated toward the anode as indicated by dotted arrow. $\Delta V_{jk}$, $\Delta L_{jk}$ and $c_{jk}$ indicated the local electric potential difference, gap and average concentration between $j$th and $k$th microelectrode. The region between two measured electrodes were denoted as red ($c_{13}$), yellow ($c_{35}$) and green ($c_{57}$) rectangles. (b) Corresponding local average concentration as a function of time for $c_{13}$, $c_{35}$ and $c_{57}$. Each region reached plateau concentration (~ 0.6 $c_0$) when the propagating fluorescent dye reached the left side of rectangle. Steady state response of plateau concentration was plotted in the inset.

**Figure 4.** (a) The schematic domain of flow field and associated concentration profile for analytical investigation. (b) Ratio of plateau concentration to bulk concentration as a function of initial bulk concentration. The plateau concentration phenomenon occurred at every concentration and its values were ~ 0.6 $c_0$.

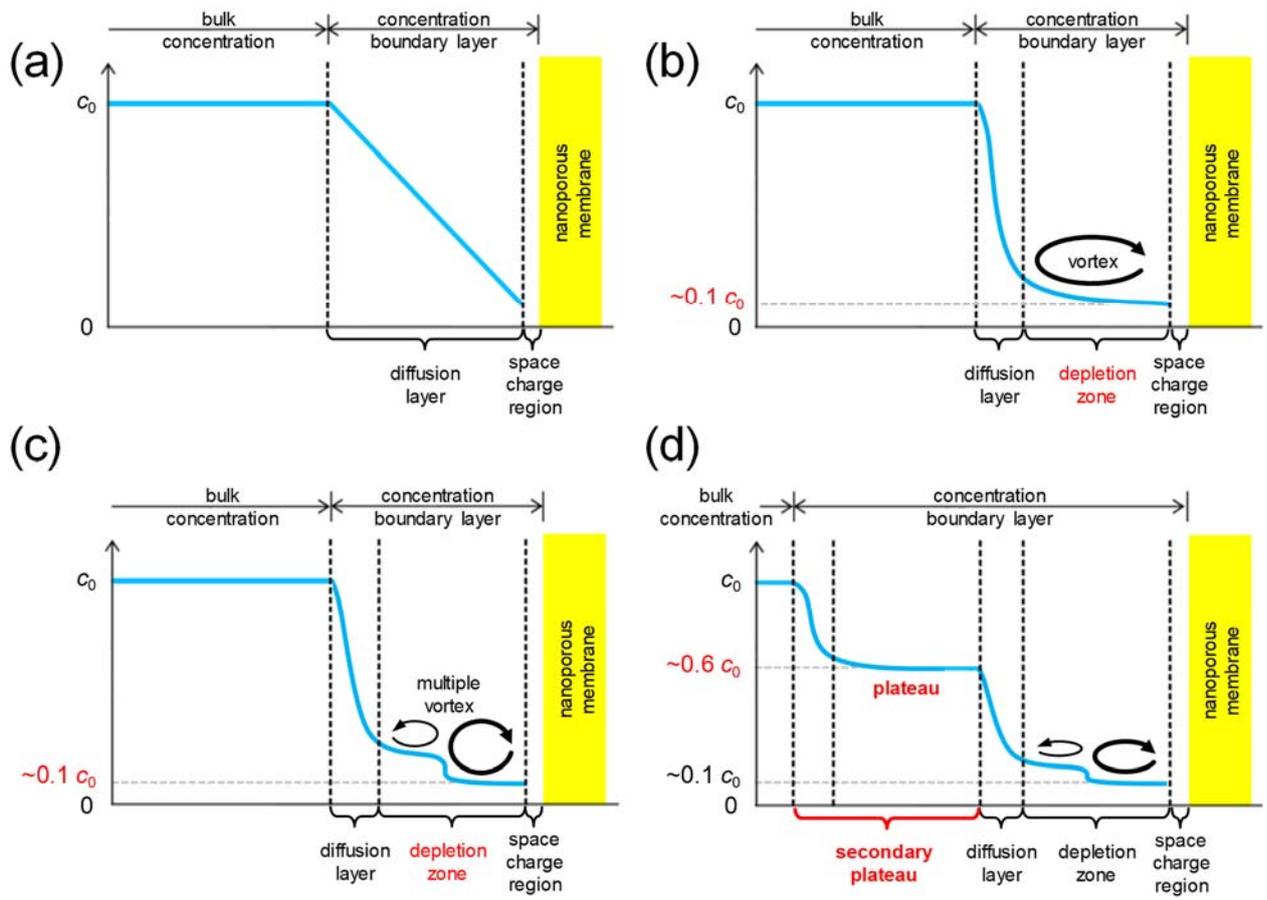

Figure 1

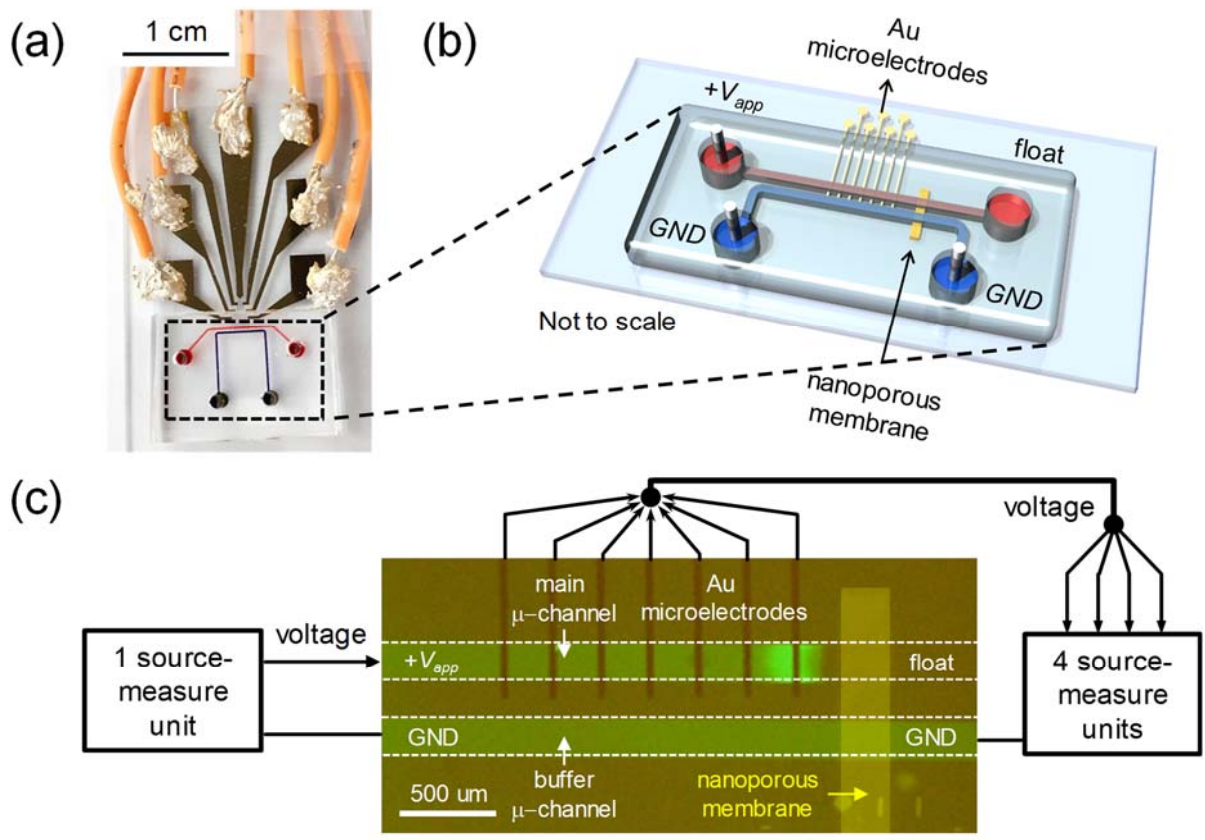

Figure 2

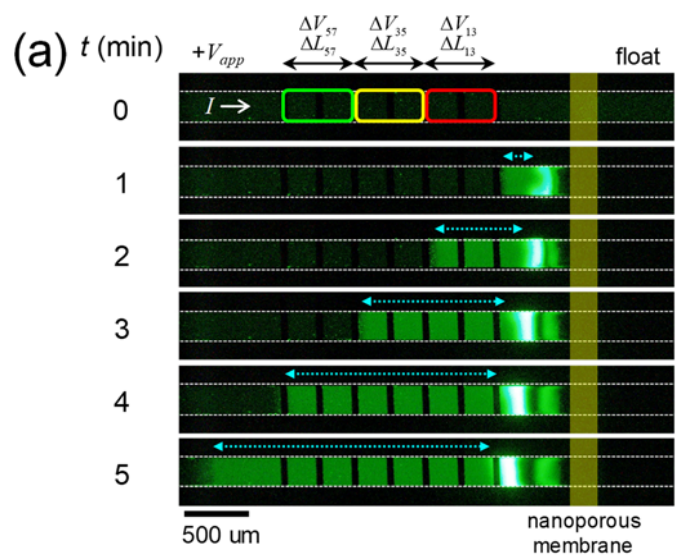

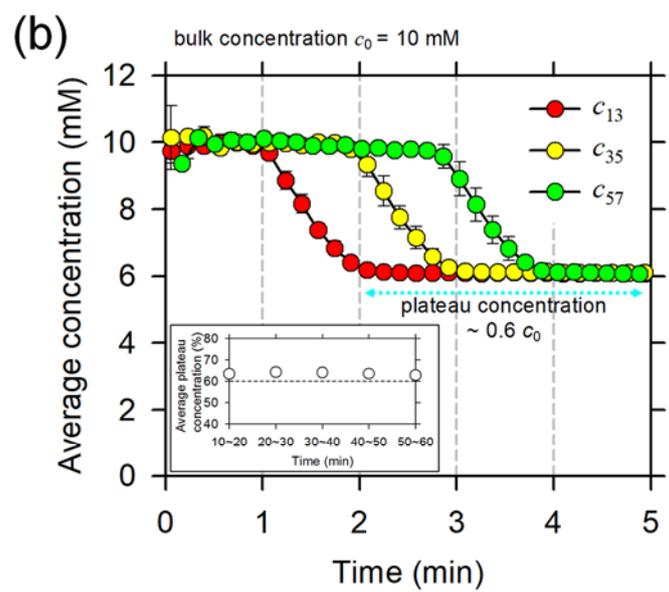

Figure 3

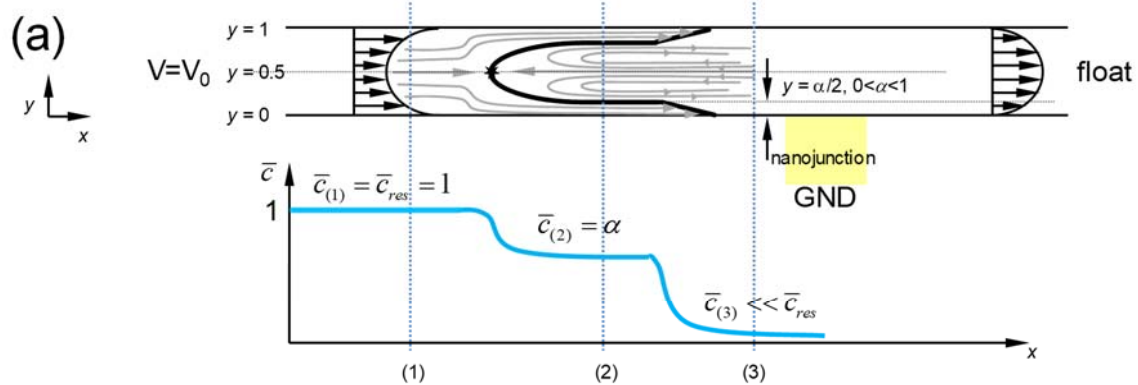

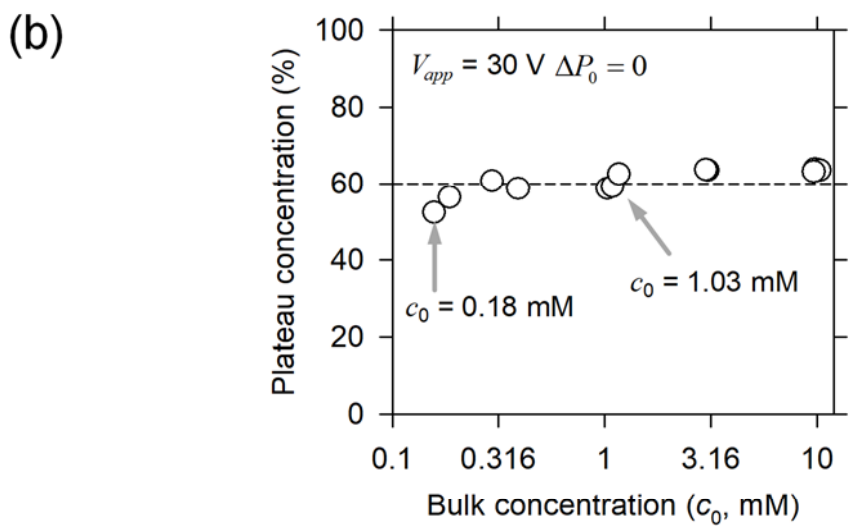

Figure 4